\documentclass[lettersize,journal]{IEEEtran}
\PassOptionsToPackage{hyphens}{url}

\usepackage{makecell}
\usepackage{longtable}
\usepackage{booktabs}
\usepackage{float}
\usepackage{amsmath,amsfonts}
\usepackage{algorithmic}
\usepackage{algorithm}
\usepackage{array}
\usepackage[caption=false,font=normalsize,labelfont=sf,textfont=sf]{subfig}
\usepackage{textcomp}
\usepackage{stfloats}
\usepackage{url}
\usepackage{verbatim}
\usepackage{graphicx}
\usepackage{cite}
\usepackage{multirow} 
\usepackage[colorlinks,linkcolor=red,anchorcolor=green,citecolor=blue]{hyperref}
\usepackage{cleveref}

\usepackage[table]{xcolor}
\crefname{figure}{fig}{figures}
\Crefname{figure}{Fig}{Figures}
\begin{document}

\title{A Trustworthy Multi-LLM Network: Challenges, Solutions, and A Use Case}

\author{Haoxiang Luo, Gang Sun,~\IEEEmembership{Senior Member,~IEEE}, Yinqiu Liu,  \\ Dusit Niyato,~\IEEEmembership{Fellow,~IEEE}, Hongfang Yu,~\IEEEmembership{Senior Member,~IEEE}, \\ Mohammed Atiquzzaman,~\IEEEmembership{Senior Member,~IEEE}, Schahram Dustdar,~\IEEEmembership{Fellow,~IEEE}
\thanks{H. Luo, G. Sun, and H. Yu are with the University of Electronic Science and Technology of China, Chengdu 611731, China (e-mail: lhx991115@163.com; \{gangsun, yuhf\} @uestc.edu.cn). Y. Liu and D. Niyato are with Nanyang Technological University, Singapore 639798 (e-mail: yinqiu001@e.ntu.edu.sg; dniyato@ntu.edu.sg). M. Atiquzzaman is with the University of Oklahoma, Norman, OK 73019 USA (e-mail: atiq@ou.edu). S. Dustdar is with the TU Wien, Vienna 1040, Austria, and also with the Universitat Pompeu Fabra, Barcelona 08002, Spain (e-mail: dustdar@dsg.tuwien.ac.at).}
\thanks{The corresponding author: Gang Sun.}
\vspace{-0.3cm}
}



\maketitle

\begin{abstract}
Large Language Models (LLMs) demonstrate strong potential across a variety of tasks in communications and networking due to their advanced reasoning capabilities. However, because different LLMs have different model structures and are trained using distinct corpora and methods, they may offer varying optimization strategies for the same network issues. Moreover, the limitations of an individual LLM's training data, aggravated by the potential maliciousness of its hosting device, can result in responses with low confidence or even bias. To address these challenges, we propose a blockchain-enabled collaborative framework that connects multiple LLMs into a Trustworthy Multi-LLM Network (MultiLLMN). This architecture enables the cooperative evaluation and selection of the most reliable and high-quality responses to complex network optimization problems. Specifically, we begin by reviewing related work and highlighting the limitations of existing LLMs in collaboration and trust, emphasizing the need for trustworthiness in LLM-based systems. We then introduce the workflow and design of the proposed Trustworthy MultiLLMN framework. Given the severity of False Base Station (FBS) attacks in B5G and 6G communication systems and the difficulty of addressing such threats through traditional modeling techniques, we present FBS defense as a case study to empirically validate the effectiveness of our approach. Finally, we outline promising future research directions in this emerging area.

\end{abstract}


\section{Introduction} \label{sec-I}

\IEEEPARstart {L}{arge} language models (LLMs) have become the cornerstone of artificial intelligence (AI), showing great potential in natural language understanding and generation tasks \cite{hu2024federated}. LLM provides services to users in the form of AI-Generated Content (AIGC), which has been widely used in various aspects of society, such as education, healthcare, and information. In particular, it is promising to use LLM to solve various optimization problems in networks, such as optimal resource or power allocation strategies \cite{zhang2024large}. In contrast to traditional Deep Learning (DL)-based approaches, LLM-based approaches can directly understand the natural language intent of the user and efficiently match its requirements. Moreover, LLM shows strong zero-shot generation ability, which can be directly applied to some optimization problems without retraining. Additionally, agentic AI (thought chain-of-thought, knowledge graph, etc.) and various resource-efficient fine-tuning techniques can further enhance the LLM's ability to perceive the environment and process complex downstream tasks. Thus, it can provide more reasonable outputs than DL methods \cite{wen2024generative}. As a result, LLM builds a new paradigm for optimizing problems in a variety of scenarios, such as 5G and 6G communications.

However, different LLMs have been developed using diverse training data and methodologies, leading to variations in their outputs for the same input. Additionally, outdated or limited training corpora can cause LLMs to generate biased or low-confidence responses, a phenomenon commonly referred to as hallucination. Furthermore, the generalization capability of a single LLM remains limited, making it difficult to adapt effectively across diverse network scenarios and wireless systems \cite{feng2024don}. To address these limitations, leveraging multiple LLMs to collaboratively respond to the same query is gaining increasing attention. For example, Wang et al. \cite{wang2025performance} employed GPT-3, GPT-4o, Llama3-8B, Llama3-Chinese, Doubao, SparkDesk, Qwen, and Kimi collectively to provide personalized care services for the elderly. Similarly, Marro et al. \cite{marro2024scalable} proposed \emph{Agora}, a scalable communication protocol designed to automate and coordinate collaboration among multiple LLMs.


Building on automated interactive communication among LLMs, the concept of a Multi-LLM Network (MultiLLMN) has been introduced to enable effective collaboration among multiple LLMs in jointly delivering intelligent services. Although promising, the deployment of MultiLLMN for network optimization in 5G, 6G, and other communication systems requires addressing two critical challenges:
\begin{itemize}
\item \textbf{Response Efficiency:} A key challenge lies in determining how to efficiently select the best response from multiple LLMs and ensure consensus among them. This is essential for enabling MultiLLMN to meet the ultra-low-latency requirements of modern communications.
\item \textbf{Response Trustworthiness:} LLMs deployed on compromised or untrusted devices may exhibit malicious behavior due to Trojans, viruses, or the intent of their operators. Moreover, intentionally harmful LLMs, such as WormGPT \cite{firdhous2023wormgpt}, can actively mislead users by generating deceptive responses. Such threats pose significant risks to the reliability of MultiLLMN and can severely degrade network performance. \end{itemize}

Therefore, we propose to use the blockchain as a solution to MultiLLMN to provide a trustworthy optimization method for the network. On the one hand, blockchain consensus enables MultiLLMN to make the best quality responses without relying on trusted third parties. On the other hand, the immutable and traceable nature of the blockchain guarantees the credibility of the response generated by MultiLLMN. To the best of our knowledge, this is the first work on blockchain-driven multiple LLMs to provide optimized services for networks. Specifically, our contributions are as follows.

\begin{itemize} 

\item To address the limitations of a single LLM, such as biases and hallucinatory outputs, as well as its limited generalizability to intricate wireless networks, we introduce MultiLLMN, a collaborative framework designed to organize multiple LLMs. It aims to enhance the accuracy and adaptability of AI-driven solutions for complex wireless communication systems.

\item To mitigate the risks posed by potentially malicious LLM behaviors and to ensure reliable content generation, we design a blockchain to support the MultiLLMN framework. This integration gives rise to the Trusted MultiLLMN, a secure and decentralized paradigm that leverages blockchain-based validation to enhance transparency, auditability, and robustness in LLM-empowered wireless network optimization.

\item To validate the feasibility and effectiveness of Trustworthy MultiLLMN, we conduct a case study on False Base Station (FBS) attacks in 5G and 6G communication systems. These attacks can lead to significant economic losses and pose serious security threats. Furthermore, as the number of FBS increases, the complexity of optimization problems grows disproportionately, making the design of effective defense mechanisms increasingly challenging. In this context, the Trusted MultiLLMN is employed to deliver precise and trustworthy responses to such attacks without relying on complex system modeling or extensive pre-training.


\end{itemize}







 
\section{The Trustworthy Multi-LLM Network} \label{sec-II}


 \subsection{Related Work of Multi-LLM and Trustworthy LLM}

 Due to the differences in the learning corpus, training path, and scenario-based orientation of different LLMs, their answers to the same questions will inherently differ. Moreover, some LLMs could have limitations and obsolescence in the training data, resulting in biased content generation and even hallucinations. Thus, researchers propose the collaboration of multiple LLMs to provide a reliable response to users. For instance, Feng et al. \cite{feng2024don} proposed a collaborative framework that integrates three LLMs. It systematically addresses knowledge gaps in the single LLM, such as those arising from incomplete or outdated training data, by leveraging cross-model validation and complementary expertise among the LLM ensemble. The framework achieved a 19.3\% performance improvement over a single LLM on four tasks with different knowledge areas, including common sense selection, word puzzles, natural language reasoning, and news summary. In addition, Owens et al. \cite{owens2024multi} considered the bias of individual LLM outputs, which is also caused by limited training data. Despite some progress in natural language processing (NLP) techniques, such as data enhancement and model fine-tuning, biased results persist. Therefore, they built a communication model of multi-LLM to reduce bias in the generated results. 

 Additionally, due to the existence of malicious LLMs such as WormGPT and the potential malicious behavior of the devices carrying LLMs, research on trustworthy LLMs has been conducted. For example, Liu et al. \cite{liu2024blockchain} used blockchain to provide trusted endorsement and protection for AIGC products. They also provide traceable verification services for altering ownership of AIGC products based on blockchain and incentive mechanisms. In addition, Luo et al. \cite{luo2024bc4llm} considered the trustworthiness of LLM from three aspects: learning corpus, training process, and generated content. Then, they also highlighted that blockchain will play an important role in these areas.

 However, the aforementioned studies still have not solved the problem of trusted collaboration among multiple LLMs. This will fundamentally address biases, hallucinations, and credibility challenges in the content generated by LLMs\cite{lu2024merge}.

\subsection{Concepts for MultiLLMN and Trustworthy MultiLLMN}

\begin{figure*}[!t]
   \centering
   \includegraphics[width=7 in]{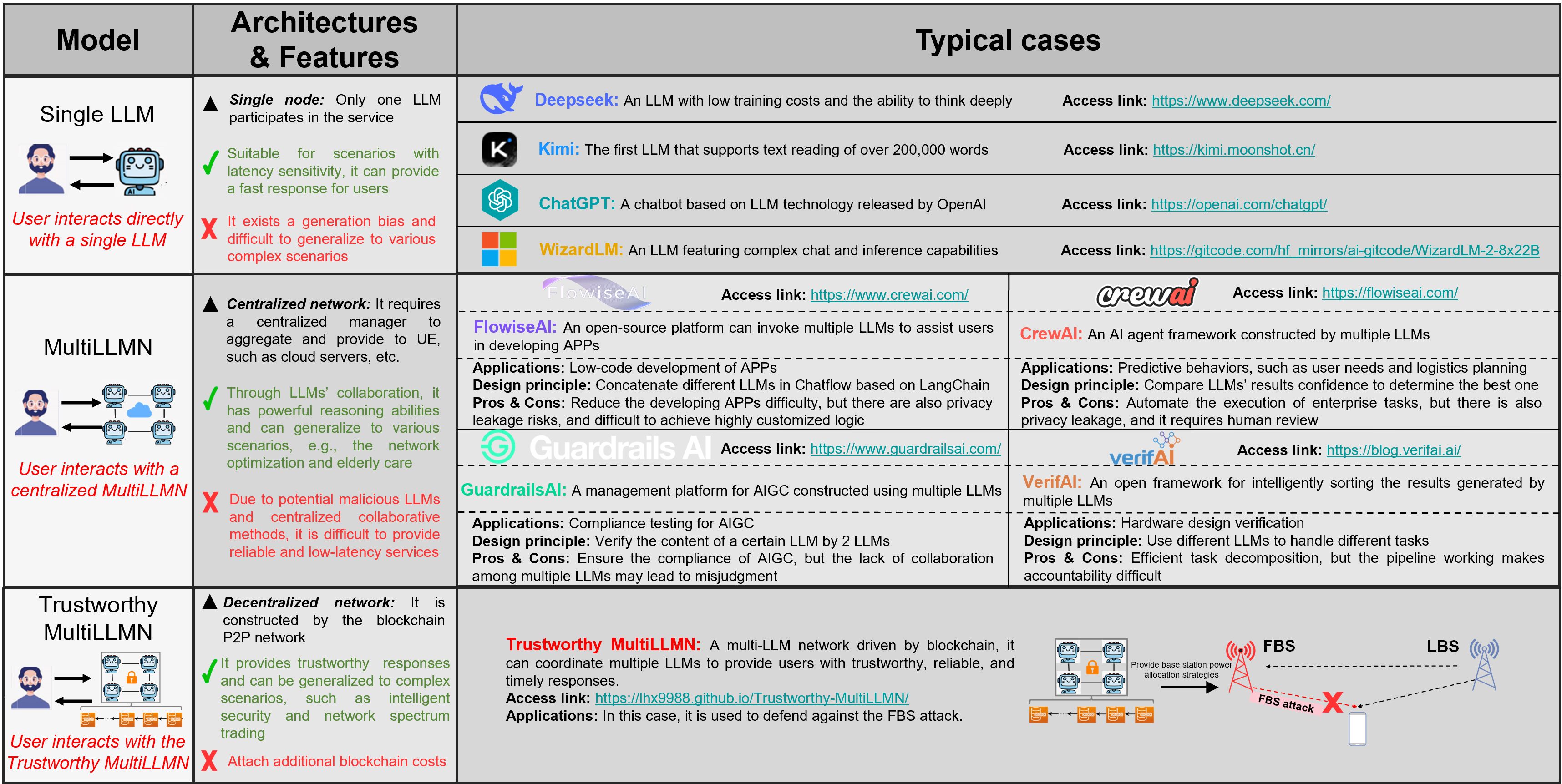}
   \caption{ Compared with a single LLM and MultiLLM, Trustworthy MultiLLMN has distributed characteristics and is less likely to be disturbed by malicious LLMs. It has higher network stability and robustness, and can provide trusted services for user.}
   \label{fig3}
\end{figure*}

Fig. \ref{fig3} illustrates the comparison between a single LLM, MultiLLMN, and Trustworthy MultiLLMN, including their architectures, application scenarios, and typical cases.

\begin{itemize} 

\item \textbf{Single LLM} uses one LLM to answer users' questions or address optimization goals. In this regard, the LLM can directly provide a timely response to the question. However, the user cannot refer to the answers from other LLMs, even if the content generated by this LLM is incomplete, biased, or hallucinatory. 

\item \textbf{MultiLLMN} is an intelligent network composed of multiple LLMs that can collaboratively respond to requests. It not only reflects the common capability of each LLM, but also avoids the outcome with bias or hallucinations from a single LLM.
The typical applications of this method are 360 AI Assistant\footnote{https://bot.360.com} and Corex\footnote{https://link.zhihu.com/?target=https\%3A//github.com/QiushiSun/Corex}. The former can invoke three LLMs to work collaboratively. The latter is designed by the Shanghai AI Lab and enables multiple LLMs to reason together. Amazon\footnote{https://aws.amazon.com/cn/blogs/machine-learning/multi-llm-routing-strategies-for-generative-ai-applications-on-aws/} has also explored message routing strategies among multiple LLMs. Other examples are shown in Fig. \ref{fig3}. However, there are also problems with response efficiency and credibility in existing approaches. Traditional centralized networks require a central node to coordinate these responses from different LLMs, which is time-consuming. The centralized network architecture also has the risk of a single point of failure. Furthermore, there exists a potential for malicious activities on the device that hosts the LLM, which could subsequently influence other benign LLMs within the MultiLLMN. This compromised interaction may lead to the dissemination of unreliable or untrusted responses.

\item \textbf{Trustworthy MultiLLMN} integrates blockchain into the MultiLLMN. Owing to the security characteristics of blockchain, it holds the potential to ensure the traceability and immutability of the generated content. Furthermore, the consensus empowers the MultiLLMN to execute dependable decision-making processes efficiently, thereby obviating the necessity for reliance on a centralized arbiter. Although the integration of blockchain effectively mitigates the challenges of response efficiency and trust, it concurrently entails supplementary costs associated with blockchain implementation, specifically encompassing storage and communication costs. 

\end{itemize}

In communication networks such as 5G and 6G, the open nature of wireless channels makes them inherently vulnerable to various security threats. Within such environments, it becomes difficult to determine whether the devices hosting LLMs exhibit malicious behavior. As a result, the generation of reliable and trustworthy content becomes critically important. The Trusted MultiLLMN framework, by advancing toward a trust-centric architecture, is well-positioned to play a pivotal role in securing LLM collaboration under these conditions.

\subsection{Trustworthy MultiLLMN Workflow}

\begin{figure}[!t]
\centering
 \includegraphics[width=3.2 in]{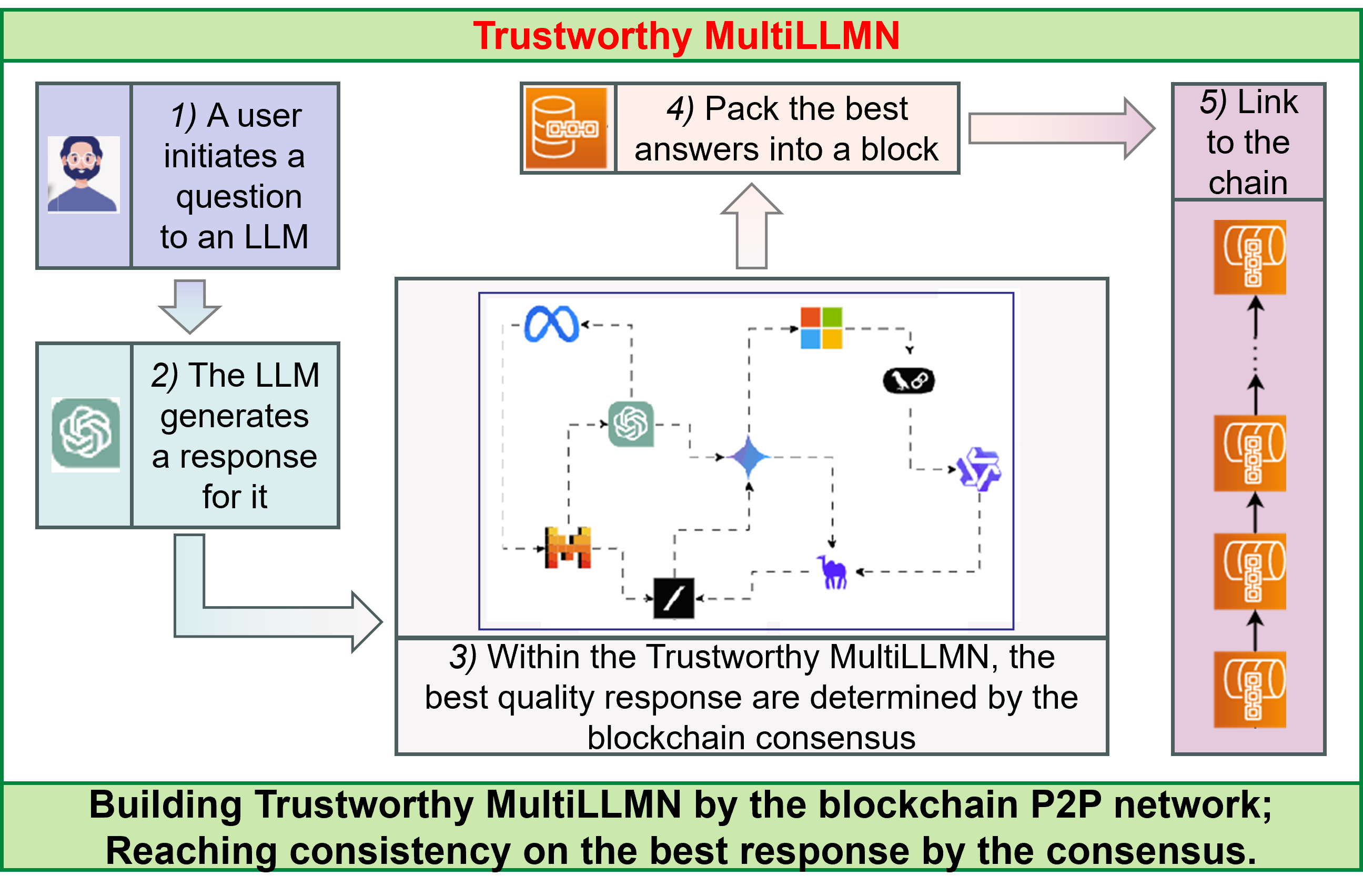}
   \caption{The blockchain-driven Trustworthy MultiLLMN. The response provided by a certain LLM to a user will be verified and compared by all LLMs in Trustworthy MultiLLMN. Thus, it ensures that the blockchain-driven network can provide the user with the highest quality and trusted answer.}
\label{fig4}
\end{figure}

In this part, we introduce how blockchain facilitates the Trustworthy MultiLLMN. Fig. \ref{fig4} shows a trusted MultiLLMN built on a blockchain Peer-to-Peer (P2P) network, including several currently mainstream LLMs, such as Llama 3, WizardLM 2, GPT-4, and Gemini 2 Flash. Regardless of their specific type, LLMs possess the ability not only to generate content but also to assess its accuracy. As a result, each LLM is regarded as one full blockchain node, contributing to the consensus process for selecting the most suitable response to network optimization problems. The workflow is outlined as follows:

\emph{1) User Requests:} A user submits a response request to Trustworthy MultiLLMN. The user can be a network operator in the scenario of defending against FBS attacks. Simply, the network operator can obtain a trusted power allocation from MultiLLMN. We will elaborate in detail in Sec-\ref{sec-III}.

\emph{2) LLM Answer Generation:} Each LLM generates a response based on the user's need. Then, LLMs interact among them via broadcast protocols in the blockchain P2P network.

\emph{3) Blockchain Consensus:} Consensus is the key to determining which of these responses from different LLMs is the optimal. We adopt the voting consensus to compare and select the final output, which is then fed back to the user by consensus leaders. In the subsequent simulation in Sec-\ref{sec-IV}, we will compare the response efficiency of different consensus mechanisms to the Trustworthy MultiLLMN.

\emph{4) Block Creation:}  The optimal output determined by blockchain consensus is packaged in a block, which each LLM confirms the block. To ensure that consensus results are secure, immutable, and traceable, the block contains the hash value of the optimal solution and its timestamp.

\emph{5) Blockchain Extension:}  Blocks containing consensus results are linked on the chain and stored in a distributed manner on smart devices running LLM.

\textbf{Lesson Learned}: By integrating blockchain, Trustworthy MultiLLMN effectively addresses the issues of output bias, illusion, and low credibility caused by data limitations and malicious behavior. The decentralized consensus replaces the centralized coordination node. It automatically coordinates multiple LLMs to generate an optimal network optimization plan (e.g., the power allocation strategy for 5G/6G anti-counterfeiting FBS attacks), avoiding a single point of failure and resisting the interference of malicious nodes. Later, we will show how it can be applied in a real network.

\begin{figure*}[!t]
   \centering
   \includegraphics[width=6 in]{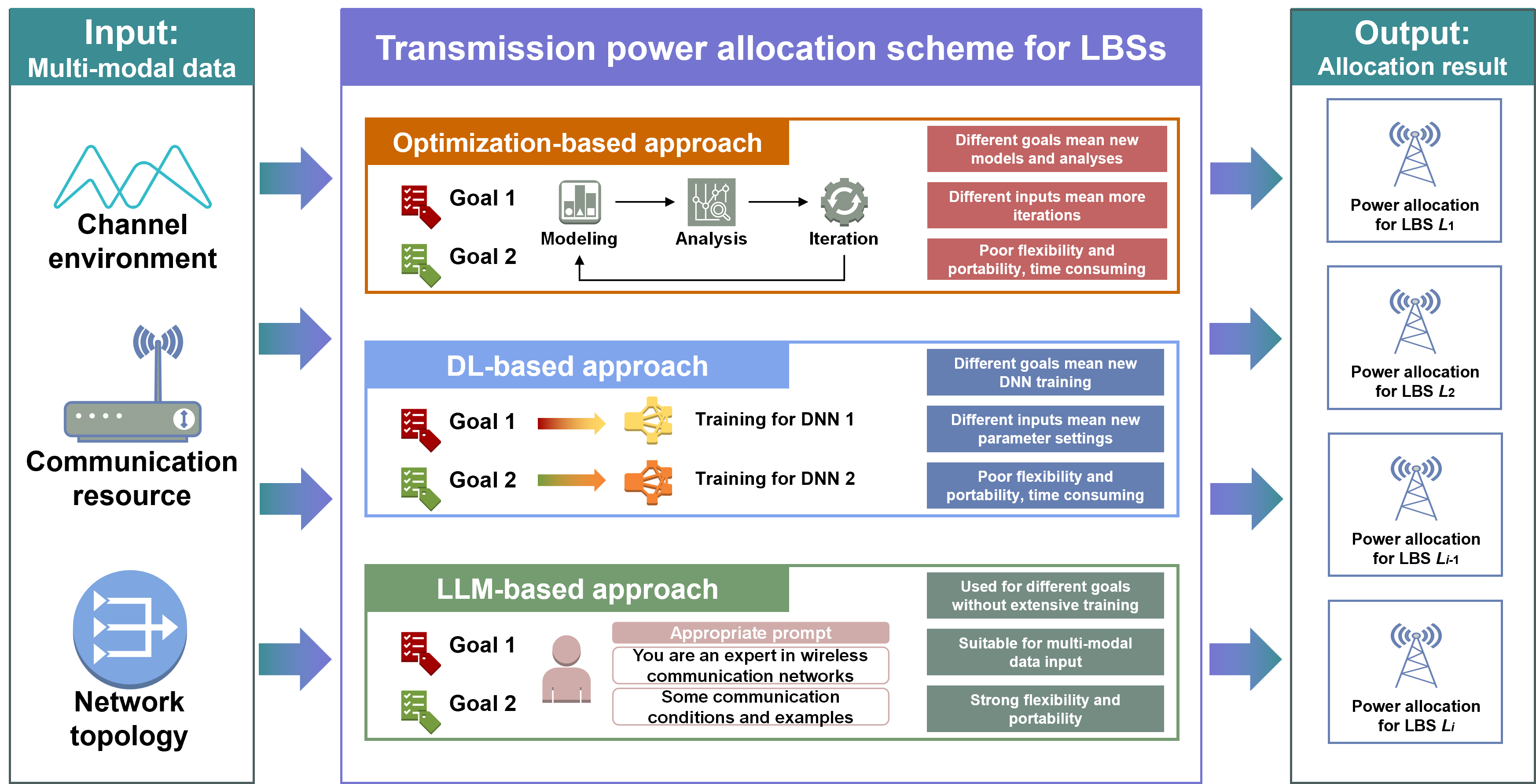}
   \caption{Different wireless transmission power distribution methods for LBSs. Both optimization and DL-based methods suffer from poor flexibility and portability, while the LLM-based method only needs appropriate prompts to generate power allocation strategies.}
   \label{fig2}
\end{figure*}

\section{Case Study: Trustworthy MultiLLMN-Enabled Defense Mechanism for FBS Attacks} \label{sec-III}

To maintain network security, we present how to use Trustworthy MultiLLMN to enable wireless communication systems against FBS attacks as a case study. It is a typical cyber attack and has caused serious economic losses to 5G communications. In January 2023, China used 923 radio monitoring vehicles, 2,457 positioning devices, and dispatched 2,459 monitoring personnel for 37,575 hours to combat FBS attacks\footnote{https://cbgc.scol.com.cn/news/4035087}. It is foreseeable that FBS attacks will also cause significant economic losses to 6G communication systems. 

\subsection{Motivations and LLM Advantages}

Before discussing the LLM-based method to defend against FBS attacks to highlight its motivation, we first introduce traditional optimization-based and DL-based methods. Fig. \ref{fig2} shows the workflow and differences among three methods.

One is the optimization-based approach, which can analytically deal with formulaic optimization problems. Optimization problems are often complex and difficult to solve, involving integer value parameters and non-convex functions. Therefore, we tend to transform and simplify the problem to make it easier to deal with. Then, the iterative method is used to find the optimal solution. Such methods often have problems such as low solving efficiency. The other is a DL-based approach, which utilizes a specially designed deep neural network (DNN) to approximate the optimal scheme. However, to do so, targeted DNNs should be trained from scratch. This often takes a considerable amount of time/resources and a lot of reliable data. In addition, their common shortcomings are the lack of flexibility and portability. Specifically, the optimization model and the DNN are designed for specific network scenarios. When the channel environment, network topology, communication resources, etc., change, we may need to reformulate the optimization model or train the DNN. This means additional and significant time costs and computing overhead.

In contrast, LLM provides a new way to implement network optimization. Since it has been pre-trained on a wide range of datasets, it is not necessary to build models for specific tasks. Only a few rounds of appropriate prompts can accurately solve various optimization problems \cite{liu2024large}. Additionally, in a complex wireless communication system, factors including multi-modal data of channel environment, communication resource, network topology, etc., often defy processing or comprehension through conventional methodologies. In this regard, LLMs offer a distinct advantage. Meanwhile, LLM has excellent reasoning ability and can often find the underlying patterns behind complex data to generate more reasonable responses than traditional methods \cite{zhang2024generative}. However, the LLM generation scheme still presents challenges in the face of high reliability and trustworthiness requirements. Especially in the FBS attack scenario, the open network environment will further introduce malicious interference to Legitimate Base Stations (LBSs) and UEs, which will affect the effectiveness of the defense scheme provided by LLM. Thus, it is necessary to use Trustworthy MultiLLMN to deal with this attack.

\subsection{Power Allocation Model for FBS Attacks}

As described in \cite{wang2024protecting}, the success of an FBS attack depends on the Signal-to-Interference-plus-Noise Ratio (SINR) between the attacking FBS and the target LBS, a metric fundamentally determined by the transmit power of both devices. Therefore, for LBSs to protect a User Equipment (UE) by preventing the UE connecting to FBSs, the wireless communication system is crucial for the transmission power distribution of each LBS under the limited total power.

We can define the power allocation of $i$ LBSs as $p_1$, $p_3$, $p_3$,..., $p_{i-1}$, $p_i$ with $\sum_{i}^n p_{i}\leq p_{total}$, where $p_{total}$ denotes the total transmission power. In general, to optimize communication performance, the total power is allocated to each LBS. Potential FBSs around LBSs, to replace LBs to bind with UE and obtain its private system information. We assume that an FBS attacks LBS $L_i$ and successfully binds to UE, the probability is $P_{FBS_i}$. Then, our optimization goal is to maximize the average probability against FBS attacks through reasonable transmit power allocation for LBSs, namely $\max \frac{\sum_{i}^n(1-P_{FBS_i})}{n}$, where $n$ is the total number of LBSs.

\begin{figure*}[!t]
   \centering
   \includegraphics[width=7.1in]{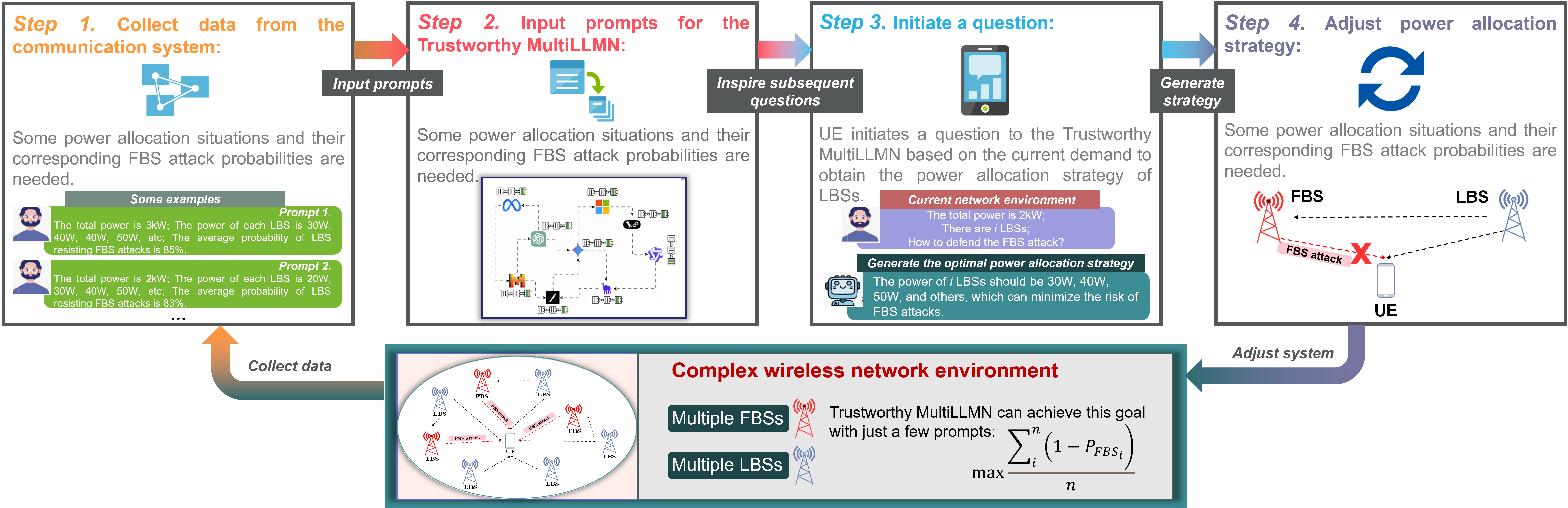}
   \caption{Trustworthy MultiLLMN-enabled defense mechanism for the FBS attack. It can provide an LBSs power allocation method for wireless communication system to resist this attack with just a few prompts.}
   \label{fig5}
\end{figure*}

\subsection{Power Allocation based on Trustworthy MultiLLMN}

To optimize the above goals under the constraint of total power, we propose a power allocation strategy based on Trustworthy MultiLLMN. This strategy-based Trustworthy MultiLLMN takes the total power and the number of LBS as input, and the power allocation of each LBS as output. Unlike traditional methods based on optimization and DL, our approach does not require a formal description of the model, but only needs a few prompts as a reference for the LLM. Therefore, the Trustworthy MultiLLMN framework can also be generalized to other communication network scenarios and optimization problems.

Specifically, the steps for the transmit power allocation strategy driven by Trustworthy MultiLLMN to resist FBS attacks are shown in Fig. \ref{fig5}.

\emph{1) Data Collection:} Collect the necessary data from the wireless network, such as historical power allocation and the corresponding FBS attack situation, as well as the total power of the system and the number of LBSs at the current stage.

\emph{2) Prompt for Trustworthy MultiLLMN:} Input the collected data into the Trustworthy MultiLLMN by a fixed format as a prompt to seek the current power allocation strategy to resist FBS attacks. The prompt is formatted as: 

\begin{itemize} 

\item \emph{``If the total power is 3kW, and the power of i LBSs is 30W, 40W, 40W, 50W, and others, then the average probability of LBSs resisting FBS attacks is 85\%"
};

\item \emph{``If the total power is 2kW, and the power of i LBSs is 20W, 30W, 40W, 50W, and others, then the average probability of LBSs resisting FBS attacks is 83\%"
}.

\end{itemize}

\emph{3) Initiate a question:} Then, in a similar format, input the current network environment to obtain the best power allocation strategy, i.e.,

\begin{itemize} 
\item  \emph{``If the total power is 2kW, what should each of i LBSs be powered to maximize the average probability of LBSs defending against FBS attacks?"}.
\end{itemize}

Trustworthy MultiLLMN provides the network operator with the optimal power allocation strategy for LBSs to effectively mitigate FBS attacks. The specific workflow of Trustworthy MultiLLMN we have described in detail in Section \ref{sec-III}-B. The output includes not only the power of each LBS, but also the average probability of an FBS attack through these LBSs. If the prompt word contains the probability of FBS completing the attack through each LBS, namely $P_{FBS_i}$, then the generated result also contains the corresponding content.

\emph{4) Adjust power allocation:} According to the output of Trustworthy MultiLLMN, the network operator further adjusts the wireless transmission power for LBSs to minimize the probability of UE accessing FBS, or accomplish the $\max \frac{\sum_{i}^n(1-P_{FBS_i})}{n}$ goal. Moreover, this strategy can serve as a new prompt for the next power allocation.

\section{Performance Evaluation} \label{sec-IV}


\subsection{Simulation Setup}

We first introduce the parameter settings. The parameters include both LLM-related and wireless communication system-related parameters.

For the former, our Trustworthy MultiLLMN contains 10 well-known LLMs, namely Llama 3.3, WizardLM 2, GPT-4o, Gemini 2 Flash, ERNIE Bot 4.0, SparkDesk V4.0, Qwen 2.5, Doubao pro 4k, HunyuanLarge, and Kimi. They will be driven by blockchain. To test the response efficiency of the blockchain-driven MultiLLMN, we set the communication environment between these LLMs to bandwidth, channel capacity, and transmission rates of 80 kHz, 15 kps, and 10 kps, respectively. For the latter, we consider a wireless network in a circle area with a radius of 5 km, where UE is at the center. In this case, we suppose there are 30 LBSs and 10 FBSs in the region. The wireless communication system has a total power of 2 kW and transmission power of 80 W per FBS. In addition, the path loss exponent is 2.5, the noise power of both UE and FBSs is $4*10^{-14}$ W, the redundancy rate is 1 bps/Hz, and the bandwidth is 20 MHz. The simulation runs with a server that contains a 96-core Intel(R) Xeon(R) Gold 5220R CPU @ 2.20GHz with 1TB of memory.

\subsection{Response Efficiency}

To verify the response efficiency of blockchain driving Trustworthy MultiLLMN, we compare several typical blockchain consensus mechanisms through simulations, including Practical Byzantine Fault Tolerance (PBFT), Trust PBFT (T-PBFT), Artificial Bee Colony PBFT (ABC-PBFT), and Votes-as-a-Proof (VaaP), which are described in detail in \cite{luo2024symbiotic}. We use the delay of these consensus-driven Trustworthy MultiLLMN operations to represent their response efficiency. 

\begin{figure}[!t]
\centering
 \includegraphics[width=2.6in]{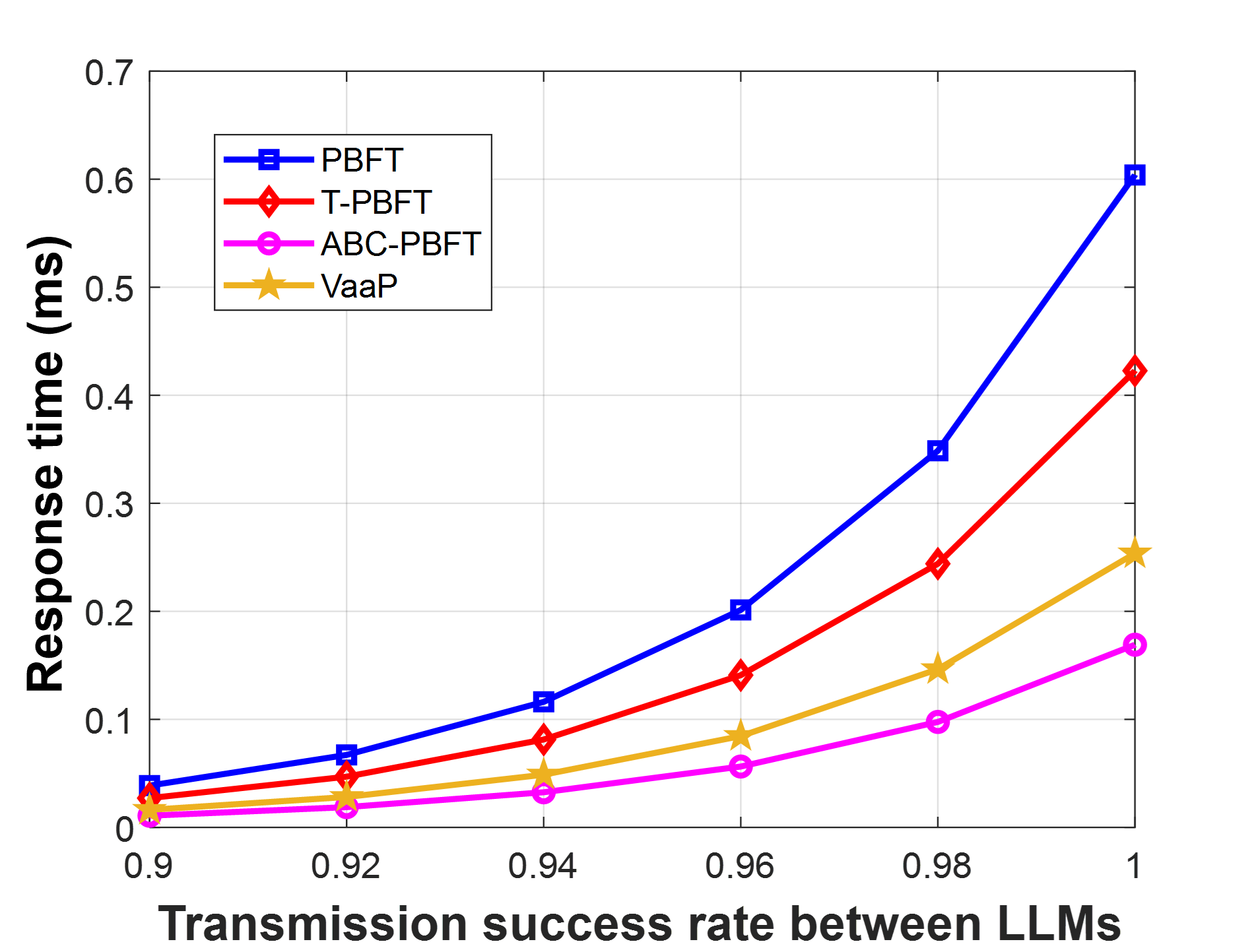}
   \caption{Response time. It compares the time required for different consensus to drive the work of Trustworthy MultiLLMN.}
\label{fig6}
\end{figure}

As shown in Fig. \ref{fig6}, it illustrates the response time (namely the latency) of the aforementioned four consensus mechanisms in driving a Trustworthy MultiLLMN. As the transmission success rate among LLMs increases, there is a corresponding rise in the response latency. This means that it is difficult to achieve both reliability and efficiency in the wireless network environment.  Notably, PBFT exhibits the highest response time, while its variant ABC-PBFT shows the lowest. T-PBFT and VaaP display intermediate levels of efficiency. The reason why ABC-PBFT has the best efficiency lies in that it screens some nodes to participate in the consensus and narrows the consensus scope. These findings offer a temporal benchmark for the operation of Trustworthy MultiLLMN in wireless environments and serve as a crucial foundation for selecting an appropriate consensus mechanism based on practical situational requirements.

\begin{figure}[!t]
\centering
 \includegraphics[width=2.6in]{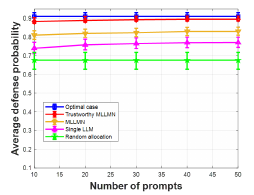 }
   \caption{Average defense probability. It proves the superiority of Trustworthy MultiLLMN in dealing with FBS attacks.}
\label{fig7}
\end{figure}

\subsection{Attack Defend Performance}

To evaluate the gain that Trustworthy MultiLLMN provides for wireless communication systems against FBS attacks, we compare it with a single LLM, MultiLLMN (without blockchain consensus participation), optimal, and random power allocation strategies. The single LLM used is ChatGPT-4o, and the PBFT consensus is used to drive Trustworthy MultiLLMN. Specifically, we calculate the average probability of all LBSs in the coverage area of the cellular network to protect the UE against FBS attacks, that is, the average resistance probability.

Fig. \ref{fig7} shows the simulation results, which clearly demonstrate the advantages of adopting consensus to establish a Trustworthy MultiLLMN framework for power allocation within wireless communication networks. Here, the MultiLLMN includes the above 10 LLMs, among which we set two LLMs to have malicious behavior. This strategy effectively mitigates FBS assaults and asymptotically approaches the theoretical optimum. Only partial prompts are needed instead of manually formulating optimization examples. The exclusion of blockchain from the MultiLLMN architecture resulted in less-than-ideal performance. It is attributed to potential malicious behavior by some LLMs that execute deceptive allocation tactics. Despite this, both MultiLLMN setups demonstrate superiority compared to a solitary LLM configuration. Crucially, any implementation of an LLM scheme markedly outperforms arbitrary power allocation methodologies, highlighting the substantial promise that LLMs hold for augmenting wireless communication systems' performance.   Moreover, the LLM-centric methodology endows each LBS with enhanced equilibrium in resisting FBS attacks.

\section{Future Directions} \label{sec-V}


\subsection{Dedicated Blockchain Consensus}

Currently, the consensus we have compared in Trustworthy MultiLLMN is not specifically tailored for it.  Consequently, both the voting modalities and the consensus processes are primed for enhancement to better suit the unique demands of MultiLLMN.  One aspect for improvement lies in the variability of response quality from individual LLMs and the trustworthiness of the devices that host them. Thus, assigning uniform voting weights to all LLMs is unwarranted. Furthermore, alongside voting-based consensus, we propose exploring a consensus model grounded in Proof-of-X, which optimizes network performance by introducing competitive elements for the privilege of responding to user queries.

\subsection{New model and system for MultiLLMN}

We use blockchain to build a Trustworthy MultiLLMN and provide services to users. The value of this concept has been proven in the use case. However, due to the complexity of the blockchain process, such as consensus and distributed storage, additional burdens are imposed on this framework. Therefore, whether some new techniques, such as the Mixture of Expert (MoE) model and the Secure Multi-Party Computation (SMPC) system, can regard each LLM as a separate expert and provide a secure collaboration mechanism for MultiLLMN is worthy of our further exploration.



\subsection{Multi-modal LLMs in MultiLLMN}

In this study, we employ a Trustworthy MultiLLMN to devise a power allocation strategy for LBSs, aimed at mitigating potential FBS security breaches. This work demonstrates the promising prospect of utilizing Trustworthy MultiLLMN to facilitate broader application prospects. In other scenarios, other types of multimodal signals, such as visual and audio, may be involved. For example, in a smart city composed of unmanned aerial vehicles and intelligent monitoring, intelligent devices equipped with the Large Vision Model (LVM) can comprehensively analyze and judge a collected image to provide reliable references for urban managers. To ensure the normal operation of MultiLLMN with multi-modal LLMs, we need to further design cross-modal collaboration strategies and consensus mechanisms.


\section{Conclusion} \label{sec-VI}

This paper presents the challenges, solutions, and use case of Trustworthy MultiLLMN based on blockchain. We have discussed that the generated results of a single LLM may be hallucinations and difficult to apply to complex network scenarios. Trustworthy MultiLLMN can avoid this critical issue and effectively mitigate bias and improve credibility. Then, we have conducted a case study to demonstrate the effectiveness and excellence of this design. In this case, Trustworthy MultiLLMN can provide a power allocation strategy for 5G or 6G communication systems against FBS attacks. Numerical results have demonstrated the advantages of our proposal over single LLM and MultiLLMN, converging notably towards the optimal solution. Furthermore, prospective avenues for further inquiry into Trustworthy MultiLLMN have been deliberated.




\bibliographystyle{IEEEtran}
\bibliography{IEEEabrv,mylib}

\begin{thebibliography}{10}
\providecommand{\url}[1]{#1}
\csname url@samestyle\endcsname
\providecommand{\newblock}{\relax}
\providecommand{\bibinfo}[2]{#2}
\providecommand{\BIBentrySTDinterwordspacing}{\spaceskip=0pt\relax}
\providecommand{\BIBentryALTinterwordstretchfactor}{4}
\providecommand{\BIBentryALTinterwordspacing}{\spaceskip=\fontdimen2\font plus
\BIBentryALTinterwordstretchfactor\fontdimen3\font minus \fontdimen4\font\relax}
\providecommand{\BIBforeignlanguage}[2]{{%
\expandafter\ifx\csname l@#1\endcsname\relax
\typeout{** WARNING: IEEEtran.bst: No hyphenation pattern has been}%
\typeout{** loaded for the language `#1'. Using the pattern for}%
\typeout{** the default language instead.}%
\else
\language=\csname l@#1\endcsname
\fi
#2}}
\providecommand{\BIBdecl}{\relax}
\BIBdecl

\bibitem{hu2024federated}
J.~Hu \emph{et~al.}, ``Federated large language model: Solutions, challenges and future directions,'' \emph{IEEE WCM}, 2024.

\bibitem{zhang2024large}
S.~Zhang \emph{et~al.}, ``Large models for aerial edges: An edge-cloud model evolution and communication paradigm,'' \emph{IEEE JSAC}, vol.~43, no.~1, pp. 21--35, 2025.

\bibitem{wen2024generative}
J.~Wen \emph{et~al.}, ``Generative {AI} for low-carbon artificial intelligence of things with large language models,'' \emph{IEEE IoTM}, vol.~8, no.~1, pp. 82--91, 2024.

\bibitem{feng2024don}
S.~Feng \emph{et~al.}, ``Don't hallucinate, abstain: Identifying {LLM} knowledge gaps via {Multi-LLM} collaboration,'' \emph{arXiv:2402.00367}, 2024.

\bibitem{wang2025performance}
S.~Wang \emph{et~al.}, ``Performance analysis on the applications of large language models: A case for elderly care,'' in \emph{2024 IEEE HPCC}.\hskip 1em plus 0.5em minus 0.4em\relax IEEE, 2024, pp. 145--151.

\bibitem{marro2024scalable}
S.~Marro \emph{et~al.}, ``A scalable communication protocol for networks of large language models,'' \emph{arXiv:2410.11905}, 2024.

\bibitem{firdhous2023wormgpt}
M.~F.~M. Firdhous \emph{et~al.}, ``{WormGPT}: A large language model chatbot for criminals,'' in \emph{2023 24th International ACIT}.\hskip 1em plus 0.5em minus 0.4em\relax IEEE, 2023, pp. 1--6.

\bibitem{owens2024multi}
D.~M. Owens \emph{et~al.}, ``A multi-{LLM} debiasing framework,'' \emph{arXiv:2409.13884}, 2024.

\bibitem{liu2024blockchain}
Y.~Liu \emph{et~al.}, ``Blockchain-empowered lifecycle management for {AI}-generated content products in edge networks,'' \emph{IEEE WCM}, vol.~31, no.~3, pp. 286--294, 2024.

\bibitem{luo2024bc4llm}
H.~Luo, J.~Luo, and A.~V. Vasilakos, ``{BC4LLM}: A perspective of trusted artificial intelligence when blockchain meets large language models,'' \emph{Neurocomputing}, vol. 599, p. 128089, 2024.

\bibitem{lu2024merge}
J.~Lu \emph{et~al.}, ``Merge, ensemble, and cooperate! a survey on collaborative strategies in the era of large language models,'' \emph{arXiv:2407.06089}, 2024.

\bibitem{liu2024large}
S.~Liu \emph{et~al.}, ``Large language models as evolutionary optimizers,'' in \emph{2024 IEEE CEC}.\hskip 1em plus 0.5em minus 0.4em\relax IEEE, 2024, pp. 1--8.

\bibitem{zhang2024generative}
R.~Zhang \emph{et~al.}, ``Generative {AI} agents with large language model for satellite networks via a mixture of experts transmission,'' \emph{IEEE JSAC}, vol.~42, no.~12, pp. 3581--3596, 2024.

\bibitem{wang2024protecting}
Z.~Wang \emph{et~al.}, ``Protecting system information from false base station attacks: A blockchain-based approach,'' \emph{IEEE TWC}, vol.~23, no.~10, pp. 13\,920--13\,934, 2024.

\bibitem{luo2024symbiotic}
H.~Luo \emph{et~al.}, ``Symbiotic blockchain consensus: Cognitive backscatter communications-enabled wireless blockchain consensus,'' \emph{IEEE/ACM ToN}, vol.~32, no.~6, pp. 5372--5387, 2024.

\end{thebibliography}



\vspace{3em}

{\footnotesize
\noindent \textbf{Haoxiang Luo} is a Ph.D. candidate at the University of Electronic Science and Technology of China (UESTC), China, and a Visiting Student at the Nanyang Technological University (NTU), Singapore.

\bigskip
\noindent \textbf{Gang Sun} (Senior Member, IEEE) is a Full Professor at the University of Electronic Science and Technology of China (UESTC), China.

\bigskip
\noindent \textbf{Yinqiu Liu} is a Ph.D. candidate at the Nanyang Technological University (NTU), Singapore.


\bigskip
\noindent \textbf {Dusit Niyato} (Fellow, IEEE) is a Full Professor at the Nanyang Technological University (NTU), Singapore. 

\bigskip

\noindent \textbf {Hongfang Yu} (Senior Member, IEEE) is a Full Professor and Vice Dean at the University of Electronic Science and Technology of China (UESTC), China.


\bigskip
\noindent \textbf {Mohammed Atiquzzaman} (Senior Member, IEEE) is an Edith Kinney Gaylord Presidential Professor at the University of Oklahoma, USA.



\bigskip
\noindent \textbf {Schahram Dustdar} (Fellow, IEEE) is a Full Professor with the Technische Universität Wien (TU Wien), Austria, and an elected member of the Academia Europaea, where he is Chairman of the Informatics Section. 




}

\vfill

\end{document}